\documentclass[preprint]{ptephy_v1}

\preprintnumber{KOBE-TH-16-02} 

\newcommand{\vev}[1]{\left\langle #1 \right\rangle}
\newcommand{\vvev}[1]{\left\langle\kern-0.3em\left\langle #1
      \right\rangle\kern-0.3em\right\rangle} 
\newcommand{\nn}{\nonumber}
\newcommand{\lb}{\left\lbrace}
\newcommand{\rb}{\right\rbrace}
\newcommand{\SL}{S_\Lambda}
\newcommand{\GL}{\Gamma_\Lambda}
\newcommand{\fy}[1]{\hbox{$#1$\kern-0.5em\raise0.3ex\hbox{/}}}
\newcommand{\Ld}[1]{\frac{\overrightarrow{\delta}}{\delta #1}}
\newcommand{\Rd}[1]{\frac{\overleftarrow{\delta}}{\delta #1}}
\newcommand{\Tr}{\mathrm{Tr}\,}
\newcommand{\N}{\mathcal{N}}

\begin{document}

\title{On the wave function renormalization \\for Wilson actions and
  their 1PI actions}


\author{Y.~Igarashi}
\author[1]{K.~Itoh}
\affil{Faculty of Education, Niigata University, Niigata 950-2181,
Japan}

\author[2]{H.~Sonoda}
\affil{Physics Department, Kobe University, Kobe 657-8501, Japan}

\begin{abstract}%
  We clarify the relation between the wave function renormalization
  for Wilson actions and that for the 1PI actions in the exact
  renormalization group formalism.  Our study depends crucially on the
  use of two independent cutoff functions for the Wilson actions.  We
  relate our results to those obtained previously by Bervillier,
  Rosten, and Osborn \& Twigg.
\end{abstract}


\maketitle

\section{Introduction\label{introduction}}

The purpose of this paper is to explain, as clearly as possible, wave
function renormalization for the Wilson actions and their 1PI
(one-particle-irreducible) actions.  Especially, we wish to elucidate
how to incorporate an anomalous dimension of the elementary field into
the Wilson and 1PI actions using the formulation of the exact
renormalization group (ERG).  We are motivated by the unsatisfactory
status of the existing literature: we cannot find any simple but
general discussion that illuminates wave function renormalization in
the ERG formalism.  We wish to expose the simple relation between the
ERG differential equations for the Wilson actions and those for the
corresponding 1PI actions with the anomalous dimension taken into
account.

Let us recall that a Wilson action comes with a momentum cutoff, say
$\Lambda$, and that its dependence on $\Lambda$ is described by the
ERG differential equation.  A 1PI action, obtained from the Wilson
action by a Legendre transformation, satisfies its own ERG
differential equation which is equivalent to that for the Wilson
action. There are many reviews available on the subject of
ERG\cite{becchi1996,morris1998,aoki2000,bagnuls2001,berges2002,gies2012,pawlowski2007,PTPreview2010,rosten2012};
we will review everything necessary in Sect.~\ref{review} to make the
paper self-contained.

An anomalous dimension $\gamma_\Lambda$ of the elementary field was
first introduced to the 1PI actions rather than the Wilson actions by
T.~Morris in \cite{morris1994,morris1994a}. (The suffix of $\gamma_\Lambda$ is for
potential $\Lambda$ dependence.)  His derivation is rather sketchy and
somewhat intuitive, but it has all the correct ingredients.  In
\cite{ellwanger1995} and \cite{berges2002}, $\gamma_\Lambda$ was
introduced differently to the 1PI actions, but the result is trivially
related to Morris's by a rescaling of the field.  It is Morris's
result, given as (\ref{ERGdiff-1PI}) in our paper, which has been
extensively used as the general form of the ERG differential equation
for the 1PI actions.

Now, what remains confusing in the literature is how to introduce
$\gamma_\Lambda$ to the Wilson actions.  Rather recently, in
\cite{osborn2011} and \cite{rosten2011}, the starting point was taken
as the ERG differential equation given earlier by Ball et
al. \cite{ball1994} for Wilson actions.  It has $\gamma_\Lambda$ as
the coefficient of a simple linear field transformation.  But this
simplicity is only apparent; the Legendre transformation, that gives
1PI actions obeying Morris's differential equation
(\ref{ERGdiff-1PI}), turns out to be very complicated.  Earlier in
\cite{bervillier2004} (and again later in \cite{bervillier2013}) C.~Bervillier
had introduced $\gamma_\Lambda$ to the Wilson actions in a way that
was later shown in \cite{bervillier2014} to admit a simple Legendre
transformation, leading to (\ref{ERGdiff-1PI}).  But the relation
between the results of \cite{osborn2011, rosten2011} and those of
\cite{bervillier2004, bervillier2013, bervillier2014} remains to be
clarified.

We will base our discussion on a recent work by one of us (H.S.)
\cite{sonoda2015} in which ERG differential equations are formulated
in terms of two separate cutoff functions, say $K_\Lambda$ and
$k_\Lambda$.  (This will be reviewed in Sect.~\ref{review}.)
$K_\Lambda$ determines the linear part of the ERG differential
equation; $k_\Lambda$, together with $K_\Lambda$, determines the
non-linear part.  The anomalous dimension is introduced via the cutoff
dependence of $K_\Lambda$.  Hence, not only the linear part but also
the non-linear part of the ERG differential equation depend on
$\gamma_\Lambda$.  It is this equation, given by
(\ref{ERGdiffwithgamma}), that we will discuss as the counterpart of
the general form (\ref{ERGdiff-1PI}) for the 1PI actions.  A simple
and well understood Legendre transformation (\ref{Legendre-SGamma})
relates the Wilson and 1PI actions.

With this understanding, the results of \cite{bervillier2004,
  bervillier2013, bervillier2014} and those of \cite{osborn2011} and
\cite{rosten2011} become transparent.  We will show that the ERG
differential equations discussed in
\cite{bervillier2004,bervillier2013,bervillier2014} and in
\cite{osborn2011, rosten2011} are obtained from the general form
(\ref{ERGdiffwithgamma}) by specific choices of our two cutoff
functions.  We relegate the technical details to appendices.  In
Appendix \ref{appendix-Bervillier}, we will show that the ERG
differential equation of \cite{bervillier2004} is obtained if the
cutoff function $k_\Lambda$ is fixed in terms of $K_\Lambda$.
Similarly, in Appendix \ref{appendix-Rosten}, we will show that the
ERG differential equation of Ball et al. \cite{ball1994} is obtained
if we choose $k_\Lambda$ specifically to remove $\gamma_\Lambda$ from
the non-linear term.  (In fact the $\gamma_\Lambda$ dependence remains
hidden in $k_\Lambda$.)  Consequently we are led to the result of
\cite{osborn2011, rosten2011} for the Legendre transformation.  In
passing we note that Wilson's original ERG differential equation and
its variation by Polchinski, both with an anomalous dimension, follow
from the general form (\ref{ERGdiffwithgamma}) by appropriate choices
of cutoff functions.  This will be explained in the second half of
Appendix \ref{appendix-supplement}.

The main part of the paper is Sect.~\ref{section-main}, where we
introduce wave function renormalization not by changing the
normalization of an elementary field but by changing that of
$K_\Lambda$.  For keeping Sect.~\ref{section-main} short and clear as
much as possible, we have collected all the relevant (mostly known but
some new) results in Sect.~\ref{review}.  Almost everything we write
in Sect.~\ref{review} has been written before, but it is handy to have
them all here.  This section also serves the purpose of making the
reader familiar with our notation.  A well-informed reader can skip
this section.  In Sect.~\ref{section-fp} we rewrite the ERG
differential equations obtained in Sect.~\ref{section-main} using the
dimensionless notation.  This is necessary to have fixed point
solutions.  We will be brief since the rewriting is pretty standard.
In Sect.~\ref{conclusion} we summarize and conclude the paper.  We
have added four appendices.  Appendix \ref{appendix-supplement}
supplements further the review given in Sect.~\ref{review}.  We hope
this makes the paper self-contained without getting too long.  We then
explain the relation of our results to those obtained by C.~Bervillier
\cite{bervillier2004, bervillier2013, bervillier2014} in Appendix
\ref{appendix-Bervillier} and those by Osborn and Twigg
\cite{osborn2011} and by Rosten \cite{rosten2011} in Appendix
\ref{appendix-Rosten}.  Finally, in Appendix \ref{appendix-examples},
we give examples of calculating the anomalous dimensions using ERG
perturbatively.

\newpage

\section{Mostly review of the relevant results\label{review}}

There are many reviews available on the subject of
ERG \cite{becchi1996,morris1998,aoki2000,bagnuls2001,berges2002,gies2012,pawlowski2007,PTPreview2010,rosten2012}.
Rather than referring the readers to them, we review all the necessary
results in this preparatory section.  Parts of 2.3-5 are new for the
viewpoint we provide.  Here is a word of warning: we have adopted a
not-so-popular convention for the sign of the Wilson action $\SL$ so
that the Boltzmann weight is $e^{\SL}$ instead of the more usual $e^{-
  \SL}$.  This reduces the number of minus signs in the formulas.  We
work in $D$ dimensional Euclidean space, and we use the notation
\begin{equation}
\int_p = \int \frac{d^D p}{(2\pi)^D},\quad
\delta (p) = (2\pi)^D \delta^{(D)} (p)
\end{equation}
for momentum integrals and the $D$-dimensional delta function.

\subsection{ERG differential equation}

Let $\SL$ be a Wilson action with a momentum cutoff $\Lambda$.  As we
lower $\Lambda$, we change $\SL$ such that physics is preserved.  
The specific $\Lambda$-dependence is given by the exact
renormalization group (ERG) differential equation
\begin{eqnarray}
- \Lambda \frac{\partial \SL [\phi]}{\partial \Lambda} &=& \int_p \Lambda
\frac{\partial \ln K_\Lambda (p)}{\partial \Lambda} \phi (p)
\frac{\delta \SL [\phi]}{\delta \phi (p)}\nn\\
&& \hspace{-1cm}+ \int_p \Lambda \frac{\partial}{\partial
      \Lambda} \ln \frac{K_\Lambda (p)^2}{k_\Lambda (p)} \cdot
    \frac{k_\Lambda (p)}{p^2} \frac{1}{2} \lb
\frac{\delta \SL[\phi]}{\delta \phi (p)} \frac{\delta
  \SL[\phi]}{\delta \phi (-p)} + \frac{\delta^2 \SL[\phi]}{\delta \phi
  (p) \delta \phi (-p)} \rb\,,\label{ERGdiff}
\end{eqnarray}
which is characterized by two positive cutoff functions $K_\Lambda
(p)$ and $k_\Lambda (p)$.\cite{sonoda2015} $K_\Lambda (p)$ must have
the properties
\begin{equation}
K_\Lambda (p) \longrightarrow \lb\begin{array}{c@{\quad}l}
\mathrm{const}& (p^2 \to 0)\,,\\
0& (p^2 \to \infty)\,,
\end{array}\right.
\end{equation}
while $k_\Lambda (p)$ must vanish as $p^2 \to 0$:
\begin{equation}
k_\Lambda (p) \stackrel{p^2 \to 0}{\longrightarrow} 0\,.
\end{equation}
Any ERG differential equation given in the past can be written as
above if the cutoff functions are chosen appropriately.  As we will
see in Sect.~\ref{section-main}, this is still the case even with an
anomalous dimension.

For example, the choice
\begin{equation}
K_\Lambda (p) = K_\Lambda^W (p) \equiv e^{- p^2/\Lambda^2},\quad
k_\Lambda (p) = k_\Lambda^W (p) \equiv \frac{p^2}{\Lambda^2}\,,
\end{equation}
was made when the ERG differential equation was written for the first
time \cite{wilson1974}. For perturbative
applications \cite{polchinski1984}, it is convenient to choose
\begin{equation}
K_\Lambda (p) = K(p/\Lambda),\quad
K (0) = 1,\quad k_\Lambda (p) = k^P_\Lambda (p) \equiv K (p/\Lambda)
\left( 1 - K (p/\Lambda)\right)\,
\end{equation}
so that $K(p/\Lambda)/p^2$ and $(1-K (p/\Lambda))/p^2$ are interpreted
as low and high momentum propagators, respectively.

\subsection{Modified correlation functions}

Using $K_\Lambda$ \& $k_\Lambda$ we can introduce modified correlation
functions as
\begin{eqnarray}
&&\vvev{\phi (p_1) \cdots \phi (p_n)}_{\SL}^{K_\Lambda, k_\Lambda}\nn\\
&&\equiv \prod_{i=1}^n \frac{1}{K_\Lambda (p_i)} \cdot
\vev{\exp \left( - \frac{1}{2} \int_p \frac{k_\Lambda (p)}{p^2}
      \frac{\delta^2}{\delta \phi (p)\delta \phi (-p)} \right) \phi
  (p_1) \cdots \phi (p_n)}_{\SL}\nn\\
&&= \prod_{i=1}^n \frac{1}{K_\Lambda (p_i)} \cdot \int [d\phi] e^{\SL
  [\phi]}\nn\\
&&\qquad\times \exp \left( - \frac{1}{2} \int_p \frac{k_\Lambda (p)}{p^2}
      \frac{\delta^2}{\delta \phi (p)\delta \phi (-p)} \right)
\, \lb\phi (p_1) \cdots \phi (p_n)\rb\,.
\label{modified}
\end{eqnarray}
$\SL$ is expected to suppress the fluctuations of high momentum modes,
and the division by $K_\Lambda (p)$ enhances those fluctuations.  The
propagator $k_\Lambda (p)/p^2$ modifies two-point functions only at
momenta of order $\Lambda$ or higher.  The modified correlation
functions thus defined are independent of the cutoff $\Lambda$ if
$\SL$ satisfies (\ref{ERGdiff}). In fact, as shown in
\cite{sonoda2015}, we can derive the ERG differential equation
(\ref{ERGdiff}) by demanding the $\Lambda$-independence of (\ref{modified}).

In the first half of Appendix \ref{appendix-supplement} we solve
(\ref{ERGdiff}) and show the $\Lambda$ independence of
(\ref{modified}).

\subsection{Equivalent Wilson actions}

Let $K'_\Lambda$ be a cutoff function alternative to $K_\Lambda$.
Substituting $\left(K_\Lambda (p)/K'_\Lambda (p)\right) \phi (p)$ for
$\phi (p)$ in the above modified correlation functions, we obtain
\begin{equation}
\prod_{i=1}^n \frac{1}{K'_\Lambda (p)} \cdot
\vev{\exp \left( - \frac{1}{2} \int_p \frac{k_\Lambda (p)}{p^2}
\frac{K'_\Lambda (p)^2}{K_\Lambda (p)^2} \frac{\delta^2}{\delta \phi
  (p) \delta \phi (-p)} \right) \phi (p_1) \cdots \phi (p_n)}_{\SL'}
\end{equation}
where $\SL' [\phi]$ is the Wilson action obtained by the substitution:
\begin{equation}
\SL' [\phi] \equiv \SL \left[ \frac{K_\Lambda (p)}{K'_\Lambda (p)}
  \phi (p)\right]\,.\label{Sprime}
\end{equation}
If we define
\begin{equation}
k'_\Lambda (p) \equiv k_\Lambda (p) \frac{K'_\Lambda (p)^2}{K_\Lambda
  (p)^2}\,,\label{kprime}
\end{equation}
then $\SL$ with $K_\Lambda, k_\Lambda$ has the same modified
correlation functions as $\SL'$ with $K'_\Lambda, k'_\Lambda$:
\begin{equation}
\vvev{\phi (p_1) \cdots \phi (p_n)}_{\SL'}^{K'_\Lambda, k'_\Lambda}
= \vvev{\phi (p_1) \cdots \phi (p_n)}_{\SL}^{K_\Lambda, k_\Lambda}\,.
\end{equation}
These are independent of $\Lambda$ if $\SL$ satisfies the ERG
differential equation (\ref{ERGdiff}) with $K_\Lambda, k_\Lambda$.
Then, $\SL'$ is guaranteed to satisfy the ERG differential equation
with $K'_\Lambda, k'_\Lambda$.  We can regard the combination $(\SL',
K'_\Lambda, k'_\Lambda)$ as equivalent to the combination $(\SL,
K_\Lambda, k_\Lambda)$: equivalent actions give the same modified
correlation functions.  

Please note that a somewhat broader definition of equivalence was
introduced in \cite{sonoda2015}: a combination $(S,K,k)$ of a Wilson
action and two cutoff functions is regarded as equivalent to
$(S',K',k')$ if they give the same modified correlation functions.
Hence, $(\SL, K_\Lambda, k_\Lambda)$ for all $\Lambda$ belongs to the
same class of equivalent actions.  (Strictly speaking, $\Lambda$
should not be smaller than the physical mass.)  We have adopted a
narrower definition so that two equivalent actions are related to each
other by a simple linear field transformation (\ref{Sprime}).

\subsection{Functional $W_\Lambda[J]$}

Given $\SL$ satisfying (\ref{ERGdiff}) with $K_\Lambda, k_\Lambda$, we
define
\begin{equation}
\tilde{S}_\Lambda [\phi] \equiv \frac{1}{2} \int_p
\frac{p^2}{k_\Lambda (p)} \phi (p) \phi (-p) + \SL [\phi]\,.
\end{equation}
This satisfies
\begin{eqnarray}
- \Lambda \frac{\partial \tilde{S}_\Lambda [\phi]}{\partial \Lambda}
&=& \int_p \Lambda \frac{\partial \ln \frac{K_\Lambda (p)}{R_\Lambda
    (p)}}{\partial \Lambda} \phi (p) \frac{\delta \tilde{S}_\Lambda
  [\phi]}{\delta \phi (p)}\nn\\
&& + \int_p \Lambda \frac{\partial R_\Lambda (p)}{\partial \Lambda}
\frac{1}{2} \frac{K_\Lambda (p)^2}{R_\Lambda (p)^2} \lb \frac{\delta
  \tilde{S}_\Lambda [\phi]}{\delta \phi 
  (p)}\frac{\delta \tilde{S}_\Lambda [\phi]}{\delta \phi (-p)}
+ \frac{\delta^2 \tilde{S}_\Lambda [\phi]}{\delta \phi (p) \delta \phi
(-p)} \rb\,,
\end{eqnarray}
where we define
\begin{equation}
R_\Lambda (p) \equiv \frac{p^2}{k_\Lambda (p)} K_\Lambda (p)^2\,.
\end{equation}
The first term can be eliminated by the change of field variables from
$\phi (p)$ to
\begin{equation}
J(p) \equiv \frac{R_\Lambda (p)}{K_\Lambda (p)} \phi (p)\,.
\end{equation}
Defining
\begin{eqnarray}
    W_\Lambda [J] &\equiv& \tilde{S}_\Lambda
    \left[ \frac{K_\Lambda (p)}{R_\Lambda (p)} J (p)\right]\\
    &=& \frac{1}{2} \int_p J(p) \frac{1}{R_\Lambda (p)} J(-p) + \SL
    \left[ \frac{K_\Lambda (p)}{R_\Lambda (p)} J (p)\right]\,, \nn
\end{eqnarray}
we obtain
\begin{equation}
- \Lambda \frac{\partial W_\Lambda [J]}{\partial \Lambda} =
\frac{1}{2} \int_p \Lambda \frac{\partial R_\Lambda (p)}{\partial \Lambda} \lb
\frac{\delta W_\Lambda[J]}{\delta J(p)} \frac{\delta W_\Lambda[J]}{\delta J(-p)} +
\frac{\delta^2 W_\Lambda[J]}{\delta J(p) \delta J(-p)}\rb \,,
\label{W-Lambda}
\end{equation}
which depends only on the single combination $R_\Lambda (p)$ of the
two cutoff functions.  This equation was first obtained by T.~Morris
in \cite{morris1994,morris1994a}.

Given $W_\Lambda [J]$ with a choice of $R_\Lambda (p)$, the
corresponding Wilson action $\SL [\phi]$ is not uniquely determined.
To specify $\SL [\phi]$ we need to specify one more cutoff function,
say $K_\Lambda (p)$.  Then, $k_\Lambda (p)$ is given by
\begin{equation}
k_\Lambda (p) = \frac{p^2}{R_\Lambda (p)} K_\Lambda (p)^2\,,
\end{equation}
and $\SL [\phi]$ is determined as
\begin{equation}
\SL [\phi] = - \frac{1}{2} \int_p \frac{p^2}{k_\Lambda (p)} \phi (p)
\phi (-p) + W_\Lambda \left[ \frac{R_\Lambda (p)}{K_\Lambda (p)} \phi
    (p) \right]\,.
\end{equation}
If we choose $K'_\Lambda (p)$ instead of $K_\Lambda (p)$, the
resulting $\SL'$ and $k'_\Lambda$ satisfy respectively (\ref{Sprime})
and (\ref{kprime}).  Hence, the pair $(W_\Lambda, R_\Lambda)$
corresponds to a class of equivalent Wilson actions $(\SL, K_\Lambda,
k_\Lambda), (\SL', K'_\Lambda, k'_\Lambda), \cdots$ all of which give
rise to the same modified correlation functions.

\subsection{Legendre transformation\label{subsection-Legendre}}

Given $W_\Lambda [J]$, we introduce a Legendre transformation:
\begin{equation}
\tilde{\Gamma}_\Lambda [\Phi] = W_\Lambda [J] - \int_p J(-p) \Phi
(p)\,,
\label{Legendre}
\end{equation}
where, for given $\Phi (p)$, we determine $J (p)$ by
\begin{equation}
\Phi (p) = \frac{\delta W_\Lambda [J]}{\delta J(-p)}\,.
\label{Phi}
\end{equation}
The inverse transformation is given by (\ref{Legendre})
where, for given $J(p)$, $\Phi (p)$ is determined by
\begin{equation}
J(p) = - \frac{\delta \tilde{\Gamma}_\Lambda [\Phi]}{\delta \Phi (-p)}\,.
\end{equation}
Let $(\SL, K_\Lambda, k_\Lambda)$ be one of the combinations corresponding
to $W_\Lambda [J]$.  We can then rewrite (\ref{Phi}) as
\begin{equation}
\Phi (p) = \frac{1}{K_\Lambda (p)} \left( \phi (p) + \frac{k_\Lambda
      (p)}{p^2} \frac{\delta \SL [\phi]}{\delta \phi (-p)} \right)\,.
\label{Phiphi}
\end{equation}
This is a composite operator corresponding to the elementary field
$\phi (p)$.  Its modified correlation functions satisfy
\begin{eqnarray}
&&\vvev{\Phi (p) \phi (p_1) \cdots \phi (p_n)}_{\SL}^{K_\Lambda,
  k_\Lambda}\nn\\
&&\equiv \prod_{i=1}^n \frac{1}{K_\Lambda (p_i)} \cdot \vev{\Phi (p)
  \exp \left( - \frac{1}{2} \int_p \frac{k_\Lambda (p)}{p^2}
      \frac{\delta^2}{\delta \phi (p) \delta \phi (-p)} \right) \lb
  \phi (p_1) \cdots \phi (p_n)\rb }_{\SL}\nn\\
&&= \vvev{\phi (p) \phi (p_1) \cdots \phi (p_n)}_{\SL}^{K_\Lambda,
  k_\Lambda}\,.
\end{eqnarray}

It is a general property of the Legendre transformation that the
second order differentials 
\begin{equation}
\frac{\delta W_\Lambda [J]}{\delta J(p) \delta J(-q)} = \frac{\delta
  \Phi (-p)}{\delta J(-q)},\quad
(-) \frac{\delta \tilde{\Gamma}_\Lambda [\Phi]}{\delta \Phi (p) \delta
  \Phi (-q)} = \frac{\delta J(-p)}{\delta \Phi (-q)}
\end{equation}
are the inverse of each other:
\begin{equation}
\int_q \frac{\delta^2 W_\Lambda [J]}{\delta J(p) \delta J(-q)}
(-) \frac{\delta^2 \tilde{\Gamma}_\Lambda [\Phi]}{\delta \Phi (q) \delta
  \Phi (-r)} = \delta (p-r)\,.\label{inverse}
\end{equation}
We will use the notation
\begin{equation}
G_{\Lambda; p,-q} [\Phi] \equiv \frac{\delta^2 W_\Lambda [J]}{\delta
  J(p) \delta J(-q)}\label{Gdef}
\end{equation}
when we prefer to regard this as a functional of $\Phi$.

Another general property of the Legendre transformation is that
$W_\Lambda [J]$ and $\tilde{\Gamma}_\Lambda [\Phi]$ share the same
$\Lambda$-dependence:
\begin{eqnarray}
- \Lambda \frac{\partial \tilde{\Gamma}_\Lambda [\Phi]}{\partial
  \Lambda} &=& - \Lambda \frac{\partial W_\Lambda [J]}{\partial
  \Lambda}\nn\\
&=& \frac{1}{2} \int_p \Lambda \frac{\partial R_\Lambda (p)}{\partial
  \Lambda} \lb \Phi (p) \Phi (-p) + \frac{\delta^2 W_\Lambda
  [J]}{\delta J(p) \delta J(-p)}\rb\nn\\
&=& \frac{1}{2} \int_p \Lambda \frac{\partial R_\Lambda (p)}{\partial
  \Lambda} \lb \Phi (p) \Phi (-p) + G_{\Lambda; p,-p} [\Phi] \rb\,,
\end{eqnarray}
where we have used (\ref{W-Lambda}) and (\ref{Gdef}).

We now define the 1PI action $\GL [\Phi]$ so that
\begin{equation}
\tilde{\Gamma}_\Lambda [\Phi] = - \frac{1}{2} \int_p R_\Lambda (p)
\Phi (p) \Phi (-p) + \GL [\Phi]\,.
\end{equation}
The excluded term is often called a scale dependent mass term.  The
1PI action has a very simple $\Lambda$-dependence:
\begin{equation}
- \Lambda \frac{\partial \GL [\Phi]}{\partial \Lambda} = \frac{1}{2}
\int_p \Lambda \frac{\partial R_\Lambda (p)}{\partial \Lambda}\,
G_{\Lambda; p,-p} [\Phi]\,.\label{Gamma-Lambda}
\end{equation}
Using $\GL$ we can rewrite (\ref{inverse}) as
\begin{equation}
\int_q G_{\Lambda; p,-q} [\Phi]
\left( R_\Lambda (q-r) \delta
    (q-r) - \frac{\delta^2 \GL [\Phi]}{\delta \Phi (q) \delta \Phi
      (-r)} \right)
= \delta (p-r)\,.
\end{equation}

It is important to note that the 1PI action $\GL [\Phi]$ (with
$R_\Lambda$) corresponds one-to-one to $W_\Lambda [J]$ (with
$R_\Lambda$).  Hence, all the equivalent combinations $(\SL,
K_\Lambda, k_\Lambda)$, giving rise to the same modified correlation
functions, correspond to the same 1PI action.  We end this section by
writing down the Legendre transformation (\ref{Legendre}) using $\SL$
instead of $W_\Lambda$:
\begin{equation}
- \frac{1}{2} \int_p R_\Lambda (p) \Phi (p) \Phi (-p) + \GL [\Phi]
= \frac{1}{2} \int_p \frac{p^2}{k_\Lambda (p)} \phi (p) \phi (-p) +
\SL [\phi] - \int_p \frac{R_\Lambda (p)}{K_\Lambda (p)} \phi (-p) \Phi
(p)\,.\label{Legendre-SGamma}
\end{equation}

\section{Wave function renormalization for the Wilson and 1PI actions\label{section-main}}

After a long preparation, we are ready to discuss wave function
renormalization in the ERG formalism.  We first introduce a cutoff
dependent positive wave function renormalization constant $Z_\Lambda$.
We denote the anomalous dimension by
\begin{equation}
\gamma_\Lambda = - \Lambda \frac{\partial}{\partial \Lambda} \ln
\sqrt{Z_\Lambda}\,.\label{gammaLambda}
\end{equation}
Physics dictates an appropriate choice of $Z_\Lambda$.  For now, we
can keep it arbitrary.  We wish to construct a Wilson action whose
modified correlation functions are proportional to appropriate powers
of $Z_\Lambda$:
\begin{equation}
\vvev{\phi (p_1) \cdots \phi (p_n)}_{\SL}^{K_\Lambda, k_\Lambda}
= Z_\Lambda^{\frac{n}{2}} \cdot
\left(\textrm{$\Lambda$-independent}\right) \,.
\label{ZLdependence}
\end{equation}
This implies
\begin{equation}
\frac{1}{Z_\Lambda^{\frac{n}{2}}} \vvev{\phi (p_1) \cdots \phi
  (p_n)}_{\SL}^{K_\Lambda, k_\Lambda} =
\vvev{\phi (p_1) \cdots \phi (p_n)}_{\SL}^{\sqrt{Z_\Lambda} K_\Lambda,
  k_\Lambda} = \left(\textrm{$\Lambda$-independent}\right)\,.
\end{equation}
Hence, $\SL$ satisfies the ERG differential equation for the cutoff
functions $K^Z_\Lambda \equiv \sqrt{Z_\Lambda}\,K_\Lambda$ and
$k_\Lambda$:
\begin{eqnarray}
- \Lambda \frac{\partial \SL[\phi]}{\partial \Lambda} &=& \int_p
\Lambda \frac{\partial \ln \left(\sqrt{Z_\Lambda} K_\Lambda
      (p)\right)}{\partial \Lambda} \phi (p) \frac{\delta \SL
  [\phi]}{\delta \phi (p)}\nn\\
&& \hspace{-1cm}+ \int_p \Lambda \frac{\partial}{\partial \Lambda} \ln
\frac{Z_\Lambda K_\Lambda (p)^2}{k_\Lambda (p)} \cdot \frac{k_\Lambda
  (p)}{p^2} \frac{1}{2} \lb \frac{\delta \SL [\phi]}{\delta \phi (p)}
\frac{\delta \SL [\phi]}{\delta \phi (-p)} + \frac{\delta^2 \SL
  [\phi]}{\delta \phi (p) \delta \phi (-p)} \rb\nn\\
&=&  \int_p
\Lambda \frac{\partial \ln K_\Lambda
      (p)}{\partial \Lambda} \phi (p) \frac{\delta \SL
  [\phi]}{\delta \phi (p)}\nn\\
&& \hspace{-1cm}+ \int_p \Lambda \frac{\partial}{\partial \Lambda} \ln
\frac{K_\Lambda (p)^2}{k_\Lambda (p)} \cdot \frac{k_\Lambda
  (p)}{p^2} \frac{1}{2} \lb \frac{\delta \SL [\phi]}{\delta \phi (p)}
\frac{\delta \SL [\phi]}{\delta \phi (-p)} + \frac{\delta^2 \SL
  [\phi]}{\delta \phi (p) \delta \phi (-p)} \rb\nn\\
&& \hspace{-1cm}- \gamma_\Lambda \int_p \left[
\phi (p) \frac{\delta \SL [\phi]}{\delta \phi (p)}
+ \frac{k_\Lambda (p)}{p^2} \lb\frac{\delta \SL [\phi]}{\delta \phi (p)}
\frac{\delta \SL [\phi]}{\delta \phi (-p)} + \frac{\delta^2 \SL
  [\phi]}{\delta \phi (p) \delta \phi (-p)} \rb\right]\,.
\label{ERGdiffwithgamma}
\end{eqnarray}
This is the general form of the ERG differential equation with an
anomalous dimension.  (This result is not unknown; its dimensionless
form was given in \cite{sonoda2015} as (43).)  

The last term proportional to $\gamma_\Lambda$ is worthy of a comment.
It is an equation-of-motion composite operator that counts the number
of $\phi$'s:
\[
\N_\Lambda [\phi] \equiv - \int_p K_\Lambda (p) e^{-\SL}
\frac{\delta}{\delta \phi (p)} \left[ \Phi (p) e^{\SL} \right]\,,
\]
where $\Phi (p)$ is defined by (\ref{Phiphi}).  Using (\ref{Phiphi}),
we obtain
\begin{eqnarray}
\N_\Lambda [\phi] &=& - \int_p e^{-\SL} \frac{\delta}{\delta \phi (p)}
\left[ \left( 
        \phi (p) + \frac{k_\Lambda (p)}{p^2} \frac{\delta \SL}{\delta
          \phi (-p)} \right) e^{\SL} \right]\nn\\
&=& - \int_p \left[ \phi (p) \frac{\delta \SL [\phi]}{\delta \phi (p)}
+ \frac{k_\Lambda (p)}{p^2} \lb\frac{\delta \SL [\phi]}{\delta \phi (p)}
\frac{\delta \SL [\phi]}{\delta \phi (-p)} + \frac{\delta^2 \SL
  [\phi]}{\delta \phi (p) \delta \phi (-p)} \rb\right] \label{Number}
\end{eqnarray}
up to an additive field independent constant.  This has the modified
correlation functions
\begin{eqnarray}
&&\vvev{\N_\Lambda \, \phi (p_1) \cdots \phi (p_n)}_{\SL}^{K_\Lambda,
  k_\Lambda}\nn\\
&\equiv& \prod_{i=1}^n \frac{1}{K_\Lambda (p_i)} \cdot
\vev{\N_\Lambda\, \exp \left(- \frac{1}{2} \int_p \frac{k_\Lambda
        (p)}{p^2} \frac{\delta^2}{\delta \phi (p) \delta \phi (-p)}
  \right) \lb\phi (p_1) \cdots \phi (p_n)\rb}_{\SL}\nn\\
&=& n\, \vvev{\phi (p_1) \cdots \phi (p_n)}_{\SL}^{K_\Lambda,
  k_\Lambda}\,.
\end{eqnarray}
$\N_\Lambda$ is a particular example of equation-of-motion composite
operators, also called redundant operators, marginal operators, or
exactly marginal redundant operators in the literature.  (See
\cite{wegner1974} for the original discussion.  In the context of ERG,
see, for example, \cite{sonoda2007}, \cite{osborn2008}, Appendix A of
\cite{osborn2011}, and the reviews \cite{PTPreview2010,rosten2012}.)

Using the prescription given in Sect.~\ref{subsection-Legendre}, we
can construct a 1PI action.  The result depends on whether we take
$K_\Lambda$ or $K^Z_\Lambda = \sqrt{Z_\Lambda} K_\Lambda$ as one of
the cutoff functions together with the fixed $k_\Lambda$.  Let us
first consider the combination $(\SL, K_\Lambda, k_\Lambda)$ that
corresponds to $Z_\Lambda$-dependent modified correlation functions.  We
then obtain
\begin{equation}
W_\Lambda [J] \equiv \tilde{S}_\Lambda \left[ \frac{K_\Lambda (p)}{R_\Lambda (p)} J
    (p) \right]
\end{equation}
and the 1PI action
\begin{equation}
- \frac{1}{2} \int_p R_\Lambda (p) \Phi (p) \Phi (-p) + \GL [\Phi] =
W_\Lambda [J] - \int_p J(-p) \Phi (p)\,.\label{Gamma1}
\end{equation}
We repeat the above for the combination $(\SL, K^Z_\Lambda,
k_\Lambda)$ that corresponds to $\Lambda$-independent modified
correlation functions.  Since
\begin{equation}
R^Z_\Lambda (p) \equiv \frac{p^2}{k_\Lambda (p)} K^Z_\Lambda (p)^2
= Z_\Lambda R_\Lambda (p)\,,
\end{equation}
we obtain
\begin{equation}
W^Z_\Lambda [J] \equiv \tilde{S}_\Lambda \left[ \frac{K^Z_\Lambda (p)}{R^Z_\Lambda
      (p)} J(p) \right] = \tilde{S}_\Lambda \left[ \frac{K_\Lambda
      (p)}{R_\Lambda (p)} \frac{J (p)}{\sqrt{Z_\Lambda}} \right] = W_\Lambda
\left[ \frac{J}{\sqrt{Z_\Lambda}} \right]
\end{equation}
and
\begin{eqnarray}
- \frac{1}{2} \int_p Z_\Lambda R_\Lambda (p) \Phi (p) \Phi (-p) +
\GL^Z [\Phi] &=& W^Z_\Lambda [J] - \int_p J(-p) \Phi (p)\nn\\
&=& W_\Lambda \left[\frac{J}{\sqrt{Z_\Lambda}}\right]
- \int_p \frac{J(-p)}{\sqrt{Z_\Lambda}} \sqrt{Z_\Lambda} \Phi (p)\,.
\end{eqnarray}
Comparing this with (\ref{Gamma1}), we obtain
\begin{equation}
\Gamma_\Lambda^Z [\Phi] = \GL \left[ \sqrt{Z_\Lambda} \,\Phi \right]\,.
\end{equation}
Please note that the same 1PI action is obtained from any combination
$(\SL', K'_\Lambda, k'_\Lambda)$ equivalent to $(\SL, K^Z_\Lambda,
k_\Lambda)$.  For example, with the choice $K'_\Lambda = K_\Lambda,
k'_\Lambda = k_\Lambda/Z_\Lambda$, the Wilson action
\begin{equation}
\SL' [\phi] = \SL \left[ \sqrt{Z_\Lambda}\, \phi \right]
\end{equation}
gives the same $\GL^Z [\Phi]$.

Let us find the $\Lambda$-dependence of the 1PI actions.  The
prescription of Sect.~\ref{subsection-Legendre} applies directly to
$\GL^Z [\Phi]$ that corresponds to $\Lambda$-independent modified
correlation functions.  We obtain
\begin{equation}
- \Lambda \frac{\partial \GL^Z [\Phi]}{\partial \Lambda} 
= \frac{1}{2} \int_p \Lambda \frac{\partial \left(Z_\Lambda R_\Lambda
      (p)\right)}{\partial \Lambda} \,G^Z_{\Lambda; p,-p} [\Phi]\,,
\end{equation}
where
\begin{equation}
\int_q G^Z_{\Lambda; p, -q} [\Phi] \left( Z_\Lambda R_\Lambda (q)
  \delta (q-r) - \frac{\delta^2 \GL^Z [\Phi]}{\delta \Phi (q) \delta
      \Phi (-r)} \right) = \delta (p-r)\,.
\end{equation}
This result for $\GL^Z$ is given in \cite{ellwanger1995} for gauge
theories and in \cite{berges2002} for generic scalar theories.
Consequently, the ERG differential equation for
\begin{equation}
\GL [\Phi] = \GL^Z \left[ \frac{\Phi}{\sqrt{Z_\Lambda}} \right]
\end{equation}
is obtained as
\begin{equation}
- \Lambda \frac{\partial \GL [\Phi]}{\partial \Lambda} =
- \gamma_\Lambda \int_p \Phi (p) \frac{\delta \GL
[\Phi]}{\delta \Phi (p)}  + \frac{1}{2} \int_p \left( \frac{\partial R_\Lambda
      (p)}{\partial \Lambda} - 2 \gamma_\Lambda R_\Lambda (p) \right)
G_{\Lambda; p,-p} [\Phi]\,,\label{ERGdiff-1PI}
\end{equation}
where
\begin{equation}
G_{\Lambda; p,-q} [\Phi] \equiv Z_\Lambda G^Z_{\Lambda; p,-q} \left[
    \frac{\Phi}{\sqrt{Z_\Lambda}} \right]
\end{equation}
satisfies
\begin{equation}
\int_q G_{\Lambda; p,-q} [\Phi] \left( R_\Lambda (q) \delta (q-r) -
    \frac{\delta^2 \GL [\Phi]}{\delta \Phi (q) \delta \Phi (-r)}\right) =
    \delta (p-r)\,.\label{GLambda}
\end{equation}
We regard (\ref{ERGdiff-1PI}) for $\GL$, first obtained by T.~Morris
in \cite{morris1994,morris1994a}, as the general form of the ERG differential
equation for the 1PI actions.

In (\ref{ERGdiff-1PI}), the term proportional to $\gamma_\Lambda$ is
given by the equation-of-motion operator
\begin{equation}
\N_\Lambda^{\mathrm{1PI}} [\Phi] \equiv
- \int_p \left( \Phi (p) \frac{\delta \GL [\Phi]}{\delta \Phi (p)} +
    R_\Lambda (p) G_{\Lambda; p,-p} [\Phi] \right)\,.
\end{equation}
This equals $\N_\Lambda [\phi]$ given by (\ref{Number}), written in
terms of $\Phi$ instead of $\phi$.

\section{Fixed point\label{section-fp}}

So far we have considered an arbitrary anomalous dimension
$\gamma_\Lambda$ and its integral $Z_\Lambda$.  Its introduction
becomes essential when we look for a fixed point of the ERG
differential equation, either for the Wilson action or the 1PI action.
For the differential equation to have a fixed point, we need to adopt
the dimensionless notation by measuring dimensionful quantities in
units of appropriate powers of the momentum cutoff $\Lambda$.
Rewriting of the ERG differential equation (\ref{ERGdiffwithgamma})
and (\ref{ERGdiff-1PI},\ref{GLambda}) is straightforward.  We
only give results here.  We introduce a logarithmic scale parameter
$t$ by
\begin{equation}
\Lambda = \mu e^{-t}\,,
\end{equation}
where $\mu$ is an arbitrary momentum scale. A different choice of
$\mu$ amounts to a constant shift of $t$.  Denoting the cutoff
functions by
\begin{equation}
K_\Lambda (p) = K_t (p/\Lambda)\,,\quad
k_\Lambda (p) = k_t (p/\Lambda)\,,
\end{equation}
we can rewrite (\ref{ERGdiffwithgamma}) as
\begin{eqnarray}
\partial_t S_t [\phi] &=& \int_p \left[ \lb \left( - \partial_t - p_\mu
    \frac{\partial}{\partial p_\mu} \right) \ln K_t (p) +
\frac{D+2}{2} \rb \phi (p) + p_\mu \frac{\partial \phi (p)}{\partial
  p_\mu} \right] 
 \frac{\delta S_t [\phi]}{\delta \phi (p)}\nn\\
&& + \int_p \left(\partial_t + p_\mu \frac{\partial}{\partial p_\mu}
\right) \ln \frac{k_t (p)}{K_t (p)^2} \cdot \frac{k_t (p)}{p^2}
\frac{1}{2} \lb \frac{\delta S_t [\phi]}{\delta \phi (p)} \frac{\delta
  S_t [\phi]}{\delta \phi (-p)} + \frac{\delta^2 S_t [\phi]}{\delta
  \phi (p) \delta \phi (-p)} \rb\nn\\
&& - \gamma_t \int_p \left[ \phi (p) \frac{\delta S_t [\phi]}{\delta \phi
  (p)} + \frac{k_t (p)}{p^2} \lb \frac{\delta S_t [\phi]}{\delta \phi
  (p)} \frac{\delta S_t [\phi]}{\delta \phi (-p)} + \frac{\delta^2 S_t
  [\phi]}{\delta \phi (p) \delta \phi (-p)} \rb\right]\,,
\end{eqnarray}
where we have denoted $\gamma_\Lambda$ as $\gamma_t$.  Similarly, we
can rewrite (\ref{ERGdiff-1PI}) as
\begin{eqnarray}
\partial_t \Gamma_t [\Phi] &=& \int_p \left( \frac{D+2}{2} + p_\mu
    \frac{\partial}{\partial p_\mu} \right) \Phi (p) \cdot
\frac{\delta \Gamma_t [\Phi]}{\delta \Phi (p)} \nn\\
&& + \int_p \left( 2 - \left(\partial_t + p_\mu
        \frac{\partial}{\partial p_\mu} \right) \ln R_t (p) \right)
\cdot R_t (p) \frac{1}{2} G_{t; p,-p} [\Phi]\nn\\
&& - \gamma_t \int_p \left( \Phi (p) \frac{\delta \Gamma_t
      [\Phi]}{\delta \Phi (p)} + R_t (p) G_{t; p,-p} [\Phi] \right)\,,
\end{eqnarray}
where
\begin{equation}
R_t (p) \equiv \frac{p^2}{k_t (p)} K_t (p)^2\,,
\end{equation}
and (\ref{GLambda}) as
\begin{equation}
\int_q G_{t;p,-q} [\Phi] \lb R_t (q) \delta (q-r) - \frac{\delta^2
  \Gamma_t [\Phi]}{\delta \Phi (q) \delta \Phi (-r)} \rb = \delta
(p-r)\,.
\end{equation}

To obtain a fixed point, we must choose $t$-independent cutoff
functions:
\begin{equation}
\lb\begin{array}{c@{~=~}l}
K_t (p) & K(p)\,,\\
k_t (p) & k (p)\,.
\end{array}\right.
\end{equation}
Then, the above ERG differential equations become simpler:
\begin{eqnarray}
\partial_t S_t [\phi] &=& \int_p \lb - p_\mu \frac{\partial}{\partial
  p_\mu} \ln K(p) + \frac{D+2}{2}  - \gamma_t + p_\mu \frac{\partial}{\partial
  p_\mu} \rb \phi (p) \cdot \frac{\delta S_t [\phi]}{\delta
  \phi (p)}\nn\\
&& \hspace{-1cm} + \int_p \lb - p_\mu \frac{\partial}{\partial p_\mu}
\ln R(p) + 2 - 2 \gamma_t \rb \frac{k(p)}{p^2} \frac{1}{2}
\lb \frac{\delta S_t[\phi]}{\delta \phi (p)} \frac{\delta S_t
  [\phi]}{\delta \phi (-p)} + \frac{\delta^2 S_t[\phi]}{\delta \phi
  (p) \delta \phi (-p)} \rb\,,
\end{eqnarray}
and
\begin{eqnarray}
\partial_t \Gamma_t [\Phi] &=& \int_p \left( \frac{D+2}{2} - \gamma_t + p_\mu
    \frac{\partial}{\partial p_\mu} \right) \Phi (p) \cdot
\frac{\delta \Gamma_t [\Phi]}{\delta \Phi (p)} \nn\\
&& + \int_p \left(  - p_\mu
        \frac{\partial}{\partial p_\mu} \ln R (p) + 2 - 2 \gamma_t\right)
\cdot R (p) \frac{1}{2} G_{t; p,-p} [\Phi]\,,
\end{eqnarray}
where
\begin{equation}
R (p) \equiv \frac{p^2}{k (p)} K (p)^2\,,
\end{equation}
and $G_{t; p,-q} [\Phi]$ is defined by
\begin{equation}
\int_q G_{t;p,-q} [\Phi] \lb R (q) \delta (q-r) - \frac{\delta^2
  \Gamma_t [\Phi]}{\delta \Phi (q) \delta \Phi (-r)} \rb = \delta
(p-r)\,.
\end{equation}
The anomalous dimension $\gamma_t$ can be chosen so as to fix a
particular term (the kinetic term, for example) in $S_t$.
Alternatively, it can be chosen as the fixed-point value $\gamma^*$ in
a neighborhood of the fixed point.

The fixed point action $S^*$ satisfies
\begin{eqnarray}
0 &=& \int_p \lb - p_\mu \frac{\partial}{\partial
  p_\mu} \ln K(p) + \frac{D+2}{2} - \gamma^* + p_\mu \frac{\partial}{\partial
  p_\mu}  \rb \phi (p) \cdot \frac{\delta S^* [\phi]}{\delta
  \phi (p)}\nn\\
&& \hspace{-1cm} + \int_p \lb - p_\mu \frac{\partial}{\partial p_\mu}
\ln R (p) + 2 - 2 \gamma^* \rb \frac{k(p)}{p^2} \frac{1}{2}
\lb \frac{\delta S^*[\phi]}{\delta \phi (p)} \frac{\delta S^*
  [\phi]}{\delta \phi (-p)} + \frac{\delta^2 S^*[\phi]}{\delta \phi
  (p) \delta \phi (-p)} \rb\,,\label{fp-S}
\end{eqnarray}
and the corresponding 1PI action $\Gamma^*$ satisfies
\begin{eqnarray}
0 &=& \int_p \left( \frac{D+2}{2} - \gamma^* + p_\mu
    \frac{\partial}{\partial p_\mu} \right) \Phi (p) \cdot
\frac{\delta \Gamma^* [\Phi]}{\delta \Phi (p)} \nn\\
&& + \int_p \left( - p_\mu
        \frac{\partial}{\partial p_\mu} \ln R (p) + 2 - 2 \gamma^* \right)
\cdot R (p) \frac{1}{2} G^*_{p,-p} [\Phi]\,,\label{fp-Gamma}
\end{eqnarray}
where
\begin{equation}
\int_q G^*_{p,-q} [\Phi] \lb R (q) \delta (q-r) - \frac{\delta^2
  \Gamma^* [\Phi]}{\delta \Phi (q) \delta \Phi (-r)} \rb = \delta
(p-r)\,.
\end{equation}
We can solve (\ref{fp-S}) and (\ref{fp-Gamma}) only for particular
choices of $\gamma^*$.

\section{Summary and conclusions\label{conclusion}}

In this paper we have made the best effort to elucidate the structure
of the exact renormalization group both for the Wilson actions and for
the 1PI actions.  Especially, we have tried to demonstrate the
simplicity of introducing an anomalous dimension to the ERG
differential equations.  

We have started with introducing classes of equivalent Wilson actions.
A Wilson action $\SL$ with a momentum cutoff $\Lambda$ is paired with
two cutoff functions of momentum: $K_\Lambda (p)$ and $k_\Lambda (p)$.
We then construct modified correlation functions (\ref{modified})
using $K_\Lambda$ and $k_\Lambda$.  A class of equivalent Wilson
actions consists of those combinations of $(\SL, K_\Lambda,
k_\Lambda)$ giving the same modified correlation functions.  The
equivalence of $(\SL, K_\Lambda, k_\Lambda)$ and $(\SL', K'_\Lambda,
k'_\Lambda)$ demands that $R_\Lambda (p) \equiv \frac{p^2}{k_\Lambda
  (p)} K_\Lambda (p)^2$ is the same as $R'_\Lambda (p) \equiv
\frac{p^2}{k'_\Lambda (p)} K'_\Lambda (p)^2$, and that $\SL$ and
$\SL'$ are related by (\ref{Sprime}).

The crux of the paper is the observation that all the equivalent
Wilson actions correspond to the same 1PI action $\GL [\Phi]$ via the
Legendre transformation (\ref{Legendre}) (equivalently
(\ref{Legendre-SGamma}) or (\ref{Gamma1})).  This correspondence is
many-to-one, since the Wilson action depends on two cutoff functions
$K_\Lambda, k_\Lambda$; the 1PI action depends only on $R_\Lambda$.

We have introduced the anomalous dimension $\gamma_\Lambda$ of the
elementary field by demanding the $\Lambda$-dependence of the modified
correlation functions as given by (\ref{ZLdependence}).  From this we
have derived the general form (\ref{ERGdiffwithgamma}) of the ERG
differential equation for the Wilson action.  We regard
(\ref{ERGdiffwithgamma}) as the counterpart of the general form
(\ref{ERGdiff-1PI}) for the 1PI action, introduced previously by
Morris \cite{morris1994,morris1994a}.

As long as two Wilson actions share the same $R_\Lambda$, we can
transform one Wilson action to another equivalent one just by changing
$K_\Lambda$.  For example, the Wilson action introduced by Bervillier
in \cite{bervillier2004} has an arbitrary cutoff function $K_\Lambda
(p)$ which he took to be the same as $R_\Lambda (p)$.  By choosing
this $R_\Lambda (p)$ appropriately (its explicit form is given in
Appendix \ref{appendix-Rosten}), we can convert Bervillier's ERG
differential equation into the one by Ball et al. given in
\cite{ball1994}.

Though we have discussed a generic scalar theory in this paper,
nothing prevents us from introducing anomalous dimensions to the
fermionic and gauge fields by extending our results.

\section*{Acknowledgment}

The work of Y.~I. and K.~I. was partially supported by the JSPS
grant-in-aid \#R2209 and \#22540270.  The work of H.~S. was partially
supported by the JSPS grant-in-aid \# 25400258.

\appendix

\section{Supplement to Sect.~\ref{review}\label{appendix-supplement}}

\subsection{Integration of the ERG differential equation and the
  modified correlation functions}

The ERG differential equation (\ref{ERGdiff}) is a generalized
diffusion equation, and it admits a Gaussian integral formula.  In
this appendix, we first solve the cutoff dependence of the Wilson
action, and then show that the modified correlation functions
(\ref{modified}) are independent of the cutoff.

To begin with, we rewrite (\ref{ERGdiff}) as
\begin{eqnarray}
- \Lambda \frac{\partial}{\partial
  \Lambda} e^{\SL \left[K_\Lambda \phi \right]}
&=& \int_p \Lambda \frac{\partial \ln \frac{p^2 K_\Lambda (p)^2}{k_\Lambda
    (p)}}{\partial \Lambda} \cdot \frac{k_\Lambda (p)}{p^2 K_\Lambda
  (p)^2} \frac{1}{2} \frac{\delta^2}{\delta \phi (p) \delta \phi (-p)}
e^{\SL \left[K_\Lambda \phi\right]}\nn\\
&=& - \int_p \Lambda \frac{\partial \frac{1}{R_\Lambda (p)}}{\partial
  \Lambda} \frac{1}{2} \frac{\delta^2}{\delta \phi 
(p) \delta \phi (-p)} e^{\SL \left[K_\Lambda \phi\right]} \,,
\end{eqnarray}
where
\begin{equation}
R_\Lambda (p) \equiv  \frac{p^2 K_\Lambda (p)^2}{k_\Lambda (p)}\,.
\end{equation}
Integrating this, we obtain a Gaussian integral formula:
\begin{equation}
e^{S_{\Lambda_2} \left[K_{\Lambda_2} \phi\right]}
= \exp \left[ \frac{1}{2} \int_p \left(\frac{1}{R_{\Lambda_2} (p)} -
        \frac{1}{R_{\Lambda_1} (p)} \right) \frac{\delta^2}{\delta
      \phi (p) \delta \phi (-p)} \right] e^{S_{\Lambda_1}
  \left[K_{\Lambda_1} \phi\right]}
\,.\label{ERG-integral}
\end{equation}

We then introduce a generating functional
\begin{eqnarray}
Z_\Lambda [J] &\equiv& \int [d\phi] \exp \left( \SL [\phi] + \int_p
    J(-p) \frac{\phi (p)}{K_\Lambda (p)} \right)\\
&=& \int [d\phi] \exp \left( \SL \left[ K_\Lambda \phi \right] + \int_p
    J(-p) \phi (p) \right)\,.\nn
\end{eqnarray}
Using (\ref{ERG-integral}), we obtain
\begin{eqnarray}
    Z_{\Lambda_2} [J] &=& \int [d\phi] \exp \left(\int_p J(-p) \phi (p)
    \right) e^{S_{\Lambda_2} \left[K_{\Lambda_2} \phi\right]}\nn\\
    &=&  \int [d\phi] \exp \left(\int_p J(-p) \phi (p)
    \right)\nn\\
&& \times \exp \left[ \frac{1}{2} \int_p \left(\frac{1}{R_{\Lambda_2} (p)} -
            \frac{1}{R_{\Lambda_1} (p)} \right) \frac{\delta^2}{\delta
          \phi (p) \delta \phi (-p)} \right] e^{S_{\Lambda_1}
      \left[K_{\Lambda_1} \phi\right]}\nn\\
&=& \int [d\phi]
    e^{S_{\Lambda_1} \left[K_{\Lambda_1} \phi\right]} \nn\\
&&\times \exp \left[ \frac{1}{2} \int_p
        \left(\frac{1}{R_{\Lambda_2} (p)} - \frac{1}{R_{\Lambda_1}
              (p)} \right) \frac{\delta^2}{\delta 
          \phi (p) \delta \phi (-p)} \right]\exp \left(\int_p J(-p) \phi (p)
    \right)\nn\\
&=& \exp \left[ \frac{1}{2} \int_p
        \left(\frac{1}{R_{\Lambda_2} (p)} - \frac{1}{R_{\Lambda_1}
              (p)} \right) J(p) J(-p) \right] Z_{\Lambda_1} [J]\,.
\end{eqnarray}
We have thus found that
\begin{eqnarray}
&&Z_\Lambda [J] \exp \left[ - \frac{1}{2} \int_p \frac{1}{R_\Lambda (p)}
    J(p) J(-p) \right]\nn\\
&&= \int [d\phi] \exp \left[ S_\Lambda [\phi] + \int_p \left( J(-p) \frac{\phi
      (p)}{K_\Lambda (p)} - \frac{1}{2} \frac{k_\Lambda (p)}{p^2}
    \frac{J(p)}{K_\Lambda (p)} \frac{J(-p)}{K_\Lambda (p)} \right) \right]
\end{eqnarray}
is independent of $\Lambda$, and it generates the modified correlation
functions defined by (\ref{modified}):
\begin{equation}
Z_\Lambda [J] \exp \left[ - \frac{1}{2} \int_p \frac{1}{R_\Lambda (p)}
    J(p) J(-p) \right] = \sum_{n=0}^\infty \frac{1}{n!}
\int_{p_1,\cdots,p_n} J(p_1) \cdots J(p_n) \vvev{\phi (p_1) \cdots
  \phi (p_n)}_{\SL}^{K_\Lambda, k_\Lambda}.
\end{equation}

\subsection{Two examples of ERG differential equations}

\subsubsection{Wilson}

When the ERG differential equation was motivated and derived for the
first time in \cite{wilson1974}, the following particular choice was
made for the cutoff functions:
\begin{equation}
\lb\begin{array}{c@{~=~}l}
K_\Lambda (p) & K_\Lambda^W (p) \equiv \exp \left( -
    \frac{p^2}{\Lambda^2} \right)\,,\\ 
k_\Lambda (p) & k_\Lambda^W (p) \equiv \frac{p^2}{\Lambda^2}\,,\\
R_\Lambda (p) & R_\Lambda^W (p) \equiv \Lambda^2 \exp \left( - 2
    \frac{p^2}{\Lambda^2} \right)\,. 
\end{array}\right.
\end{equation}
Correspondingly, (\ref{ERGdiffwithgamma}), (\ref{ERGdiff-1PI}), and
(\ref{GLambda}) become
\begin{eqnarray}
- \Lambda \frac{\partial \SL}{\partial \Lambda} &=& \int_p \left( 2
    \frac{p^2}{\Lambda^2} - \gamma_\Lambda \right) \phi \frac{\delta
  \SL}{\delta \phi} \nn\\
&&\qquad + \int_p \left(1 + 2 \frac{p^2}{\Lambda^2} -
    \gamma_\Lambda \right) \frac{1}{\Lambda^2} \lb \frac{\delta
  \SL}{\delta \phi} \frac{\delta \SL}{\delta \phi} + \frac{\delta^2
  \SL}{\delta \phi \delta \phi} \rb,\label{ERGdiff-Wilson}\\
- \Lambda \frac{\partial \GL}{\partial \Lambda} &=& - \gamma_\Lambda
\int_p \Phi \frac{\delta \GL}{\delta \Phi} + \int_p \left(1 -
    \gamma_\Lambda + 2 \frac{p^2}{\Lambda^2} \right) \Lambda^2 e^{-
  \frac{2 p^2}{\Lambda^2}}\, G_{\Lambda; p,-p}\,,
\end{eqnarray}
and
\begin{equation}
\int_q G_{\Lambda; p,-q} \left( \Lambda^2 e^{-
      \frac{2 q^2}{\Lambda^2}} \delta (q-r) - \frac{\delta^2 \GL}{\delta
      \Phi (q) \delta \Phi (-r)} \right) = \delta (p-r)\,.
\end{equation}

\subsubsection{Polchinski\label{appendix-polchinski}}

The following choice was made in \cite{polchinski1984}: 
\begin{equation}
\lb\begin{array}{c@{~=~}l}
K_\Lambda (p) & K(p/\Lambda)\,,\\
k_\Lambda (p) & k^P _\Lambda (p) \equiv K (p/\Lambda) \left(1 -
    K(p/\Lambda) \right)\,,\\ 
R_\Lambda (p) & p^2 \frac{K(p/\Lambda)}{1 - K(p/\Lambda)}\,,
\end{array}\right.
\end{equation}
where $K(p)$ satisfies
\begin{equation}
K (p) \longrightarrow \lb\begin{array}{c@{\quad}l}
1& (p \to 0)\,,\\
0& (p \to \infty)\,.
\end{array}\right.
\end{equation}
(Only the massless case is considered here for simplicity.)  Regarding
\begin{equation}
S_{F\Lambda} [\phi] = - \frac{1}{2} \int_p \frac{p^2}{K(p/\Lambda)}
\phi (p) \phi (-p)
\end{equation}
as the free part of the Wilson action, we can develop perturbation theory.

With the above cutoff functions, (\ref{ERGdiffwithgamma}) and
(\ref{ERGdiff-1PI}) become
\begin{eqnarray}
- \Lambda \frac{\partial \SL}{\partial \Lambda} &=& \int_p
\left(\frac{\Delta (p/\Lambda)}{K(p/\Lambda)} - \gamma_\Lambda \right)
\phi \frac{\delta \SL}{\delta \phi} \nn\\
&&\qquad + \int_p \frac{1}{p^2} \left(
    \Delta (p/\Lambda) - 2 \gamma_\Lambda K(1-K) \right) \frac{1}{2}
\lb \frac{\delta
  \SL}{\delta \phi} \frac{\delta \SL}{\delta \phi} + \frac{\delta^2
  \SL}{\delta \phi \delta \phi}\rb\,,\label{ERGdiff-Polchinski}\\
- \Lambda \frac{\partial \GL}{\partial \Lambda} &=& - \gamma_\Lambda
\int_p \Phi \frac{\delta \GL}{\delta \Phi} + \frac{1}{2} \int_p
\frac{p^2}{(1-K)^2} \left( \Delta - 2 \gamma_\Lambda K(1-K) \right)
G_{\Lambda; p,-p}\,,
\end{eqnarray}
where $\Delta (p/\Lambda) = \Lambda \partial_\Lambda K(p/\Lambda)$,
and (\ref{GLambda}) becomes
\begin{equation}
\int_p G_{\Lambda; p,-q} \left( q^2
    \frac{K(q/\Lambda)}{1-K(q/\Lambda)} \delta (q-r) - 
    \frac{\delta^2 \GL}{\delta \Phi (q) \delta \Phi (-r)} \right) =
\delta (p-r)\,.
\end{equation}
(\ref{ERGdiff-Polchinski}) was derived in \cite{PTPreview2010}.

To make $\SL$ equivalent to the action satisfying the Wilson
convention, we must choose 
\begin{equation}
K(p/\Lambda) = \frac{1}{1 + \frac{p^2}{\Lambda^2} \exp \left( \frac{2
        p^2}{\Lambda^2} \right)}
\end{equation}
so that $R_\Lambda (p) = \Lambda^2 \exp (- 2 p^2/\Lambda^2)$.
We then find
\begin{equation}
\SL^W [ \phi] \equiv \SL \left[ \frac{e^{\frac{p^2}{\Lambda^2}}}{1 +
      \frac{p^2}{\Lambda^2} e^{\frac{2 p^2}{\Lambda^2}}} \phi\right]
\end{equation}
satisfies (\ref{ERGdiff-Wilson}).\cite{sonoda2015}

\section{Comments on the results of Bervillier\label{appendix-Bervillier}}

In refs. \cite{bervillier2004, bervillier2013, bervillier2014}
Bervillier gives ERG differential equations (for both Wilson and 1PI
actions) with an anomalous dimension.  He has only one arbitrary
cutoff function, since he makes a choice
\begin{equation}
\lb\begin{array}{c@{~=~}l}
K_\Lambda (p) & \tilde{P} (p/\Lambda) = K(p/\Lambda)\\
k_\Lambda (p) & \frac{p^2}{\Lambda^2} \tilde{P} (p/\Lambda) =
\frac{p^2}{\Lambda^2} K(p/\Lambda)
\end{array}\right.
\end{equation}
so that
\begin{equation}
R_\Lambda (p) = \Lambda^2 \tilde{P} (p/\Lambda) = \Lambda^2
K(p/\Lambda)\,.
\label{Bervillier-R}
\end{equation}
Using the dimensionless notation, we can write Bervillier's ERG differential
equations as
\begin{eqnarray}
\partial_t S_t [\phi] &=& \int_p \lb - p_\mu \frac{\partial \ln
  K(p)}{\partial p_\mu} + \frac{D+2}{2} - \gamma_t + p_\mu
\frac{\partial}{\partial p_\mu} \rb \phi (p) \cdot \frac{\delta S_t
  [\phi]}{\delta \phi (p)} \nn\\
&& \hspace{-1cm} + \int_p \lb (2 - 2 \gamma_t) K(p)  - p_\mu
\frac{\partial K(p)}{\partial p_\mu} \rb \frac{1}{2} \lb \frac{\delta
  S_t [\phi]}{\delta \phi (p)} \frac{\delta S_t [\phi]}{\delta \phi
  (-p)} + \frac{\delta^2 S_t [\phi]}{\delta \phi (p) \delta \phi (-p)}
\rb\,,
\end{eqnarray}
and
\begin{eqnarray}
\partial_t \Gamma_t [\Phi] &=& \int_p \left(\frac{D+2}{2}-\gamma_t +
    p_\mu \frac{\partial}{\partial p_\mu} \right) \Phi (p) \cdot \frac{\delta
      \Gamma_t [\Phi]}{\delta \Phi (p)}\nn\\
&& + \int_p \left( (2-2\gamma_t) K(p) - p_\mu \frac{\partial
      K(p)}{\partial p_\mu} \right) \frac{1}{2} G_{t;p,-p} [\Phi]\,,
\end{eqnarray}
where
\begin{equation}
\int_q G_{t;p,-q} [\Phi] \lb K(q) \delta (q-r) - \frac{\delta^2
  \Gamma_t [\Phi]}{\delta \Phi (q) \delta \Phi (-r)} \rb = \delta
(p-r)\,.
\end{equation}
(\ref{Bervillier-R}) simplifies considerably the Legendre
transformation from $S_t$ to $\Gamma_t$:
\begin{equation}
- \frac{1}{2} \int_p K(p) \Phi (p) \Phi (-p) + \Gamma_t [\Phi]
= \frac{1}{2} \int_p \frac{1}{K(p)} \phi (p) \phi (-p) + S_t [\phi] -
\int_p \phi(-p) \Phi (p)\,.
\end{equation}

To make Bervillier's Wilson action equivalent to Wilson's, we must
choose
\begin{equation}
K(p/\Lambda) = \exp \left( - \frac{2 p^2}{\Lambda^2} \right)
\end{equation}
so that $R_\Lambda (p) = \Lambda^2 \exp \left(- 2
    p^2/\Lambda^2\right)$.  We then find
\begin{equation}
\SL^W [\phi] = \SL \left[ \frac{K(p/\Lambda)}{\exp (-p^2/\Lambda^2)}
    \phi \right] = \SL \left[ \exp (-p^2/\Lambda^2) \phi \right]
\end{equation}
satisfies (\ref{ERGdiff-Wilson}), as is explained in \cite{bervillier2004}.

\section{Comments on the results of Rosten and those of Osborn and Twigg\label{appendix-Rosten}}

Osborn and Twigg \cite{osborn2011} and independently Rosten
\cite{rosten2011} have considered the ERG differential equation of
Ball et al. \cite{ball1994} given by
\begin{eqnarray}
- \Lambda \frac{\partial}{\partial \Lambda} \SL [\phi] &=&
\int_p \left(\frac{\Delta (p/\Lambda)}{K(p/\Lambda)} - \gamma_\Lambda
\right) \phi (p) \frac{\delta \SL}{\delta \phi (p)}\nn\\
&& + \int_p \frac{\Delta (p/\Lambda)}{p^2} \frac{1}{2} \lb \frac{\delta
  \SL}{\delta \phi (p)} \frac{\delta \SL}{\delta \phi (-p)} +
\frac{\delta^2\SL}{\delta \phi (p) \delta \phi (-p)} \rb\,,\label{ERGdiff-Ball}
\end{eqnarray}
where the dimensionful notation is used, and
\begin{equation}
\Delta (p/\Lambda) \equiv \Lambda \frac{\partial}{\partial \Lambda}
K(p/\Lambda)\,.
\end{equation}
They succeeded in constructing a Legendre transformation that gives a 1PI
action satisfying the ERG differential equation of the type
(\ref{ERGdiff-1PI}).  We wish to reproduce their result using our line
of reasoning.

Comparing (\ref{ERGdiff-Ball}) with (\ref{ERGdiff-Polchinski}), we
notice that (\ref{ERGdiff-Ball}) is simpler since it is missing second
order derivative terms proportional to $\gamma_\Lambda$.  But the
simplicity of (\ref{ERGdiff-Ball}) is misleading; if we try to obtain
the two cutoff functions satisfying
\begin{equation}
\lb\begin{array}{c@{~=~}l}
\Lambda \frac{\partial \ln \sqrt{Z_\Lambda} K_\Lambda (p)}{\partial \Lambda} &
\frac{\Delta (p/\Lambda)}{K(p/\Lambda)} - \gamma_\Lambda\,,\\
\Lambda \frac{\partial}{\partial \Lambda} \ln \frac{Z_\Lambda K_\Lambda
  (p)^2}{k_\Lambda (p)} \cdot k_\Lambda (p) & \Delta (p/\Lambda)\,,
\end{array}\right.
\end{equation}
so that (\ref{ERGdiff-Ball}) coincides with (\ref{ERGdiffwithgamma}),
we find the following rather complicated solutions (obtained in
\cite{rosten2011} and \cite{osborn2011}):
\begin{equation}
\lb\begin{array}{c@{~=~}l}
K_\Lambda (p) & K(p/\Lambda) \,b (p)\,,\\
k_\Lambda (p) & Z_\Lambda K(p/\Lambda)^2 \left(\int_\Lambda^{\mu}
\frac{d\Lambda'}{\Lambda'} \frac{1}{Z_{\Lambda'}} \frac{\Delta
  (p/\Lambda')}{K(p/\Lambda')^2} + a(p) \right)\,,
\end{array}\right.
\end{equation}
where $a (p), b(p)$ are arbitrary dimensionless functions of $p^2$,
independent of $\Lambda$.  ($a(p)$ must be of order $p^2/\Lambda^2$
near zero, and $b(p)$ is positive.  The $\mu$ dependence of the
integral can be absorbed by $a(p)$.)  Hence, we obtain
\begin{equation}
R_\Lambda (p) = p^2  \frac{b(p)^2}{Z_\Lambda \left( \int_\Lambda^{\Lambda_0}
\frac{d\Lambda'}{\Lambda'} \frac{1}{Z_{\Lambda'}} \frac{\Delta
  (p/\Lambda')}{K(p/\Lambda')^2} + a(p)\right)}\,.
\end{equation}
The Legendre transformation (\ref{Legendre-SGamma}) gives the 1PI
action as
\begin{equation}
\GL [\Phi] = \SL [\phi] + \frac{1}{2} \int_p R_\Lambda (p) \Phi (p) \Phi (-p) +
\frac{1}{2} \int_p \frac{p^2}{k_\Lambda (p)} \phi (p) \phi (-p) -
\int_p \frac{R_\Lambda (p)}{K_\Lambda (p)} \phi (-p) \Phi (p)\,,
\end{equation}
which satisfies (\ref{ERGdiff-1PI}) and (\ref{GLambda}) with the
anomalous dimension.

For easier comparisons with the results of
\cite{rosten2011,osborn2011}, let us use the notation of
\cite{rosten2011} to express the above cutoff functions.  We define
\begin{equation}
\sigma_\Lambda (p) \equiv Z_\Lambda K(p/\Lambda) \left(
    \int_\Lambda^\mu \frac{d\Lambda'}{\Lambda'} \frac{1}{Z_{\Lambda'}}
    \frac{\Delta (p/\Lambda')}{K(p/\Lambda')^2} + a (p) \right)\,.
\end{equation}
We then obtain
\begin{eqnarray}
\mathcal{P}_\Lambda (p) &\equiv& \frac{R_\Lambda (p)}{K_\Lambda (p)} =
p^2 \frac{b(p)}{\sigma_\Lambda (p)}\,,\\
\mathcal{Q}_\Lambda (p) &\equiv& p^2 \left(\frac{1}{k_\Lambda (p)} -
    \frac{1}{K(p/\Lambda)}\right) = \frac{p^2}{K(p/\Lambda)}
    \left(\frac{1}{\sigma_\Lambda (p)} - 1 \right)\,,\\
\mathcal{R}_\Lambda (p) &\equiv& R_\Lambda (p) = p^2
\frac{K(p/\Lambda)}{\sigma_\Lambda (p)} b(p)^2\,,
\end{eqnarray}
corresponding to Rosten's $c (p) = 0$.  Hence, denoting the
interaction part by
\begin{equation}
S_{I\Lambda} [\phi] = \SL [\phi] + \frac{1}{2} \int_p
\frac{p^2}{K(p/\Lambda)} \phi (p) \phi (-p)\,,
\end{equation}
we can rewrite the above Legendre transformation as
\begin{eqnarray}
&&\GL [\Phi] = S_{I\Lambda} [\phi] \nn\\
&& + \int_p \left( -  \mathcal{P}_\Lambda (p)
\phi (-p) \Phi (p) + \frac{1}{2}
\mathcal{Q}_\Lambda (p) \phi (p) \phi (-p) + \frac{1}{2} \mathcal{R}_\Lambda (p)
\Phi (p) \Phi (-p)\right)\,,
\end{eqnarray}
agreeing with (17) of \cite{rosten2011}.

The dimensionless ERG differential equation for the 1PI action is
obtained as
\begin{eqnarray}
0 &=& \int_p \left(\frac{D+2}{2} - \gamma^* + p_\mu
    \frac{\partial}{\partial p_\mu} \right) \Phi (p) \cdot \frac{\delta
      \Gamma^* [\Phi]}{\delta \Phi (p)}\nn\\
&& + \int_p \left( - p_\mu \frac{\partial \ln R^* (p)}{\partial p_\mu}
    + 2 - 2 \gamma^* \right) R^* (p) \frac{1}{2} G^*_{p,-p} [\Phi]
\end{eqnarray}
and
\begin{equation}
\int_q G^*_{p,-q} [\Phi] \lb R^* (q) \delta (q-r) - \frac{\delta^2
  \Gamma^* [\Phi]}{\delta \Phi (q) \delta \Phi (-r)} \rb = \delta (p-r)\,,
\end{equation}
where
\begin{equation}
\frac{1}{R^* (p)} \equiv \frac{1}{b(0)^2} \frac{1}{p^2} \int_0^\infty dt\,
e^{2 \gamma t} \frac{\Delta (p e^{-t})}{K(p e^{-t})^2}\,.
\end{equation}

\section{Perturbative examples\label{appendix-examples}}

We would like to sketch two perturbative calculations of the anomalous
dimensions using ERG.  We will use the Polchinski convention
\cite{polchinski1984} of Sect.~\ref{appendix-polchinski}:
\begin{equation}
K_\Lambda (p) = K \left(\frac{p}{\Lambda}\right)\,,\quad
k_\Lambda (p) = K \left(\frac{p}{\Lambda}\right)\left(1 - K
    \left(\frac{p}{\Lambda}\right)\right) \,.
\end{equation}
We do not need to specify an explicit form of $K$ for the 1-loop
calculations.  Denoting
\begin{equation}
\Delta \left(\frac{p}{\Lambda}\right) \equiv \Lambda
\frac{\partial}{\partial \Lambda} K \left(\frac{p}{\Lambda}\right)\,,
\end{equation}
we obtain
\begin{equation}
    \int_p \frac{1}{p^D} \Delta \left(\frac{p}{\Lambda}\right)
    \, \left(1 - K \left(\frac{p}{\Lambda}\right)\right)^n =
\frac{\Omega_{D-1}}{(2 \pi)^D} \int_0^\infty dp \frac{1}{n+1}
\frac{d}{dp} \left(1 - K(p)\right)^{n+1}
= \frac{\Omega_{D-1}}{(2 \pi)^D} \frac{1}{n+1}\,,
\end{equation}
where $\Omega_{D-1}$ is the volume of the unit $D-1$ sphere given by
\begin{equation}
\Omega_{D-1} \equiv \frac{2 \pi^{\frac{D}{2}}}{\Gamma
  \left(\frac{D}{2}\right)}\,.
\end{equation}

\subsection{$\phi^3$ theory in $D=6$}

Let us consider a real scalar theory in $D=6$ whose classical action
is given by
\begin{equation}
S_{cl} = - \int d^6 x\, \left( \frac{1}{2} \left(\partial_\mu
        \phi\right)^2 + \frac{g}{3!} \phi^3 \right)\,.
\end{equation}
Let us denote the two-point vertex function of $\SL$ by
\begin{equation}
\int_p A_2 (p) \frac{1}{2} \phi (p) \phi (-p)\,.
\end{equation}
At 1-loop, (\ref{ERGdiffwithgamma}) gives
\begin{equation}
- \Lambda \frac{\partial A_2 (p)}{\partial \Lambda} =
2 \gamma_\Lambda p^2 + 
\frac{g^2}{2} \int_q \frac{\Delta (q/\Lambda)}{q^2} \frac{1-K
  \left(\frac{p+q}{\Lambda}\right)}{(p+q)^2}\,.
\end{equation}
We determine $\gamma_\Lambda$ so that the coefficient of $p^2$ vanishes.
Expanding the above to order $p^2$, we obtain
\begin{equation}
\gamma_\Lambda = \frac{g^2}{4} \int_q \frac{\Delta (q)}{q^6}
\frac{2}{3} (1 - K(q)) = \frac{g^2}{(4 \pi)^3} \frac{1}{12}\,.
\end{equation}

\subsection{Yukawa theory in $D=4$}

We next consider a theory in $D=4$ with a real scalar and a Dirac
field.  The classical action is given by
\begin{equation}
S_{cl} = - \int d^4 x\, \left( \bar{\psi} \frac{1}{i} \fy{\partial}
    \psi + \frac{1}{2} \left(\partial_\mu \phi\right)^2 + i g \phi
    \bar{\psi} \psi \right)\,.
\end{equation}
The generalization of the ERG differential equation
(\ref{ERGdiffwithgamma}) for this case is obtained as
\begin{eqnarray}
&&- \Lambda \frac{\partial \SL [\phi, \psi, \bar{\psi}]}{\partial
  \Lambda} = \int_p \left( \Lambda \frac{\partial \ln K_\Lambda
      (p)}{\partial \Lambda} - \gamma_B \right) \phi (p) \frac{\delta
  \SL}{\delta \phi (p)}\nn\\
&& \qquad + \int_p \left( \Lambda \frac{\partial \ln K_\Lambda
      (p)}{\partial \Lambda} - \gamma_F \right) \left( \SL
    \Rd{\psi (p)} \psi (p) + \bar{\psi} (-p) \Ld{\bar{\psi} (-p)}
    \SL \right)\nn\\
&& \quad+ \int_p \left( \Lambda \frac{\partial \ln R_\Lambda
      (p)}{\partial \Lambda} - 2 \gamma_B \right) \frac{k_\Lambda
  (p)}{p^2} \frac{1}{2} \lb \frac{\delta \SL}{\delta \phi (p)}
\frac{\delta \SL}{\delta \phi (-p)} + \frac{\delta^2
  \SL}{\delta \phi (p) \delta \phi (-p)} \rb\\
&&+ \int_p \left( \Lambda \frac{\partial \ln R_\Lambda
      (p)}{\partial \Lambda} - 2 \gamma_F \right) k_\Lambda (p)
\lb \SL \Rd{\psi (p)} \frac{1}{\fy{p}} \Ld{\bar{\psi} (-p)} \SL
- \Tr \frac{1}{\fy{p}} \Ld{\bar{\psi} (-p)} \SL \Rd{\psi (p)} \rb\,,\nn
\end{eqnarray}
where $\gamma_B, \gamma_F$ are the $\Lambda$-dependent anomalous
dimensions of the scalar and Dirac fields, respectively.  For
simplicity, we have used the same cutoff functions for the scalar and
Dirac fields.

Let us denote the two-point vertex functions of $\SL$ by
\begin{equation}
\int_p \left( A_B (p) \frac{1}{2} \phi (p) \phi (-p)
+ \bar{\psi} (-p) A_F (p) \psi (p) \right)\,.
\end{equation}
At 1-loop, we obtain
\begin{eqnarray}
- \Lambda \frac{\partial}{\partial \Lambda} A_B (p) &=& 2 \gamma_B p^2
 + g^2 \int_q \Tr \frac{1}{\fy{q}} \frac{1}{\fy{q}+\fy{p}}
\lb \Delta \left(\frac{p+q}{\Lambda}\right) \left(1 -
        K\left(\frac{q}{\Lambda} \right)\right)\right.\nn\\
&&\quad\left. + \left(1 -
        K\left(\frac{p+q}{\Lambda}\right)\right) \Delta
    \left(\frac{q}{\Lambda}\right) \rb,\\
- \Lambda \frac{\partial}{\partial \Lambda} A_F (p) &=& 2 \gamma_F i
\fy{p}
 - g^2 \int_q \frac{1}{q^2 \left(\fy{q}+\fy{p}\right)} \lb
 \Delta \left(\frac{p+q}{\Lambda}\right) \left(1 -
        K\left(\frac{q}{\Lambda} \right)\right) \right.\nn\\
&&\quad\left.+ \left(1 -
        K\left(\frac{p+q}{\Lambda}\right)\right) \Delta
    \left(\frac{q}{\Lambda}\right) \rb.
\end{eqnarray}
We choose $\gamma_{B,F}$ to cancel the kinetic terms.  We then obtain
\begin{eqnarray}
\gamma_B &=& 2 g^2 \int_q \frac{\Delta (q) \left(1-K(q)\right)}{q^4} =
 2 \frac{g^2}{(4 \pi)^2}\,,\\
\gamma_F &=& \frac{g^2}{2} \int_q \frac{\Delta (q) \left(1-K(q)\right)}{q^4}
=  \frac{1}{2}
\frac{g^2}{(4 \pi)^2}\,.
\end{eqnarray}

\end{document}